\documentclass[aps,prl,showpacs,twocolumn,preprintnumbers,amsmath]{revtex4}
\usepackage{bm}
\usepackage{amsmath}
\usepackage[dvips]{graphicx}
\newlength{\textwidthm}

\makeatother

\begin{document}

\title{Comment on ``BCS superconductivity of Dirac fermions in graphene
layers''}

\author{Bruno Uchoa and A. H. Castro Neto }

\affiliation{Physics Department, Boston University, 590 Commonwealth Ave., Boston,
MA 02215 }


\maketitle
In reference\cite{kopnin}, Kopnin and Sonin (KS) apply the standard
BCS model for a two dimensional electron gas with the spectrum of
Dirac fermions, namely $\xi_{\mathbf{p}}^{\alpha}=\alpha p-\mu$,
($\alpha=\pm)$, where $\mathbf{p}$ is the momentum around the Dirac
point and $\mu$ is the chemical potential. Their attempt is a generic
derivation of superconducting properties disregarding microscopic
details and the sublattice structure in graphene. In this comment
we argue that apart from their derivation of the charge current, the
thermodynamic results of ref.\cite{kopnin} are not new and were
derived before in ref.\cite{castro neto,uchoa}; second, we show
that the spectroscopic results in ref.\cite{kopnin} such as the
superfluid velocity are inconsistent with a Hamiltonian of Dirac fermions.
Finally, we show that in spite of the fact that their final result
for the current coincides with the correct result for graphene at
$\mu=0$, the derivation for Dirac fermions requires regularization,
which is only provided by the inclusion of a periodic spectrum in
the Hamiltonian.

Ref.\cite{kopnin} starts from the usual BCS spectrum for $s$-wave
pairing, $E_{\mathbf{p}}^{\alpha}=\sqrt{(\xi_{p}^{\alpha})^{2}+|\Delta|^{2}}$
($\alpha=\pm)$, which is the same spectrum derived in Ref.\cite{castro neto,uchoa}
from a particular model of Dirac fermion superconductivity. The fact
that KS do not specify a Hamiltonian, however, does not make their
thermodynamic results more general than those previous derivations
(as claimed by them) for a trivial reason: since the free energy at
the mean field level is defined only by the spectrum\cite{FreeEnergy},
any class of BCS fermionic Hamiltonians which share the same spectrum
will have \emph{exactly} the same thermodynamic properties. Since
KS start from the same BCS spectrum as ref\cite{castro neto,uchoa},
they \emph{should necessarily} obtain the same results for the gap
equation and the critical temperature, disregarding any details of
the matrix structure in the Hamiltonian. KS describe the results in
Eq. (3)$-$(11) and the subsequent equation as if they corresponded
to a new derivation, which is not the case\cite{equations}. 

In the second part of ref.\cite{kopnin}, KS calculate the supercurrent,
$\mathbf{j}$, induced by a uniform flow of the condensate with constant
momentum $\mathbf{k}_{s}=\nabla\chi$, where $\chi=\nabla\chi\cdot\mathbf{r}$
is the phase of the superconductor order parameter, $\Delta=|\Delta|\mbox{e}^{i\chi}$.
At the charge neutrality point ($\mu=0$) the Bogoliubov-DeGennes
(BdG) equations for a \emph{Dirac} Hamiltonian with $s$-wave pairing
are\cite{BG} \begin{equation}
(\mathbf{p}+\mathbf{k}_{s})\cdot\vec{\sigma}\hat{u}+\Delta\hat{v}=E\hat{u},\quad-(\mathbf{p}-\mathbf{k}_{s})\cdot\vec{\sigma}\hat{v}+\Delta^{*}\hat{u}=E\hat{v},\label{BdGEq}\end{equation}
instead of Eq.(2) in ref.\cite{kopnin}, where $\vec{\sigma}$ are
$x,y$ Pauli matrices. These equations result in a different set of
eigenvectors and also in a different spectrum, $\sqrt{E_{\mathbf{p}}^{2}+k_{s}^{2}\pm2\sqrt{(\mathbf{p}\cdot\mathbf{k}_{s})^{2}+k_{s}^{2}|\Delta|^{2}}}$,
with distinct spectroscopic properties for finite $\mathbf{k}_{s}$.
We note that due to particle-hole symmetry, the group velocity of
the quasiparticles around the Dirac point is zero, whereas the particle-hole
charge current is finite\cite{quasiparticles}. This symmetry argument
shows that the spectrum derived in ref.\cite{kopnin} (which gives
a finite superfluid velocity at half filling) is inconsistent with
\emph{any} BdG Hamiltonian of Dirac fermions, and therefore is not
applicable to graphene.

Finally, using a covariant momentum in the BdG Hamiltonian, $\hat{H}$,
namely $\mathbf{k}_{s}=\nabla\chi-\mathbf{A}$, (or \textbf{$\mathbf{k}_{s}=-\mathbf{A}$}
by a gauge choice), where $\mathbf{A}$ is the vector potential, the
current follows from $\mathbf{j}=-\partial\langle\hat{H}\rangle/\partial\mathbf{A}$.
From Eq. (1) one finds $\mathbf{j}\propto\{D-2|\Delta|\mbox{tanh}[|\Delta|/(2T)]\}\mathbf{A}$,
where $D\gg|\Delta|$ is the band width, which accounts for the orbital
paramagnetic response of the lower band electrons, overwhelming the
diamagnetism. This term is absent from the current definition of\cite{kopnin}
without justification\cite{current2}. The diamagnetism is recovered
only if one includes the full spectrum, $\xi_{\mathbf{p}}^{\alpha}=\alpha|\phi_{\mathbf{p}}|=\alpha|\phi_{\mathbf{p}}^{\prime}+i\phi_{\mathbf{p}}^{\prime\prime}|$,
where $\phi_{\mathbf{p}}$ is a periodic function. In that case, the
graphene BdG equations, $(\phi_{\mathbf{p}-\mathbf{A}}^{\prime}\sigma_{x}+\phi_{\mathbf{p}-\mathbf{A}}^{\prime\prime}\sigma_{y})\hat{u}+\Delta\hat{v}=E\hat{u}$,
and $-(\phi_{\mathbf{p}+\mathbf{A}}^{\prime}\sigma_{x}+\phi_{\mathbf{p}+\mathbf{A}}^{\prime\prime}\sigma_{y})\hat{v}+\Delta^{*}\hat{u}=E\hat{v}$,
will give $j_{i}\propto\{S_{i}+|\Delta|^{2}\sum_{\mathbf{p}}|\partial_{p_{i}}\phi_{\mathbf{p}}|^{2}E_{\mathbf{p}}^{-1}\partial_{E_{\mathbf{p}}}[\mbox{tanh}(E_{\mathbf{p}}/2T)/E_{\mathbf{p}}]\}A_{i}$,
$(i=$$x,y$ directions) where $S_{i}$ is a surface term which is
regularized by the Brillouin zone\cite{S}.


\begin{thebibliography}{1}
\bibitem{FreeEnergy} The free energy for Dirac fermions, $\Omega$,
is shown in Eq. (24) of ref.\cite{uchoa}. The gap equation follows
from $\partial\Omega/\partial|\Delta|=0$.
\bibitem{equations}For the gap, Eq. (4), (5), (6) in ref.\cite{kopnin}
correspond to Eq. (17), (15) and (20) in ref.\cite{uchoa}. For the
critical temperature, Eq. (7), (11) and the subsequent equation in\cite{kopnin}
correspond to Eq. (A.1), and (23) in ref.\cite{uchoa}. Fig.2 in
ref.\cite{kopnin} is equivalent to Fig 5 in ref.\cite{uchoa}.
Eq. (5) and (9) in ref.\cite{kopnin} were also derived in ref.\cite{castro neto}.
\bibitem{S} $S_{i}=-\sum_{\mathbf{p},\alpha}\partial_{p_{i}}\!\left\{ \alpha(\partial_{p_{i}}E_{\mathbf{p}})n(\alpha E_{\mathbf{p}})\right\} $
is exactly zero when calculated in the whole Brillouin zone, and is
$\propto D$ for a linear (non-periodic) dispersion.
\end{thebibliography}
\end{document}